\documentclass[10pt]{article}
\usepackage{amssymb}
\usepackage{amsmath}
\usepackage{graphicx}
\usepackage{fancyhdr}
\usepackage{cite}

\begin{document}

\newcommand{\bin}[2]{\left(\begin{array}{c}\!#1\!\\\!#2\!\end{array}\right)}

\newcommand{\threej}[6]{\left(\begin{array}{ccc} #1 & #2 & #3 \\ #4 & #5 & #6\end{array}\right)}

\huge

\begin{center}
Group theory and the link between expectation values of powers of $r$ and Clebsch-Gordan coefficients
\end{center}

\vspace{0.5cm}

\large

\begin{center}
Jean-Christophe Pain$^{a,b,}$\footnote{jean-christophe.pain@cea.fr}
\end{center}

\normalsize

\begin{center}
$^a$CEA, DAM, DIF, F-91297 Arpajon, France\\
$^b$Universit\'e Paris-Saclay, CEA, Laboratoire Mati\`ere en Conditions Extr\^emes,
91680 Bruy\`eres-le-Ch\^atel, France
\end{center}

\vspace{0.5cm}

\begin{abstract}
In a recent paper [J.-C. Pain, Opt. Spectrosc. {\bf 218}, 1105-1109 (2020)], we discussed the link between expectation values of powers of $r$ and Clebsch-Gordan coefficients. In this short note we provide additional information, reminding that such a connection is a direct consequence of group theory. The hydrogenic radial wavefunctions form bases for infinite dimensional representations of the algebra of the non-compact group $O(2,1)$ and the expectation values $r^p$ and $r^{-p}$ ($p$ being positive) transform as tensors with respect to this algebra. As shown a long time ago by Armstrong [L. Armstrong Jr., J . Phys. (Paris) Suppl. C 4 {\bf 31}, 17 (1970)], analysis of matrix elements of $r^p$ and $r^{-p}$ reveals that the Wigner-Eckart theorem is valid for this group and that the corresponding Clebsch-Gordan coefficients are proportional to the usual $SO(3)$ Clebsch-Gordan coefficients. This proportionality provides simple explanations of the selection rules for hydrogenic radial matrix elements pointed out by Pasternack and Sternheimer, and the proportionality of hydrogenic expectation values of $r^p$ and $r^{-p}$ to $3jm$ symbols.
\end{abstract}

\vspace{1cm}


We choose the notation of the Clebsch-Gordan coefficient $C_{j_1m_1,~j_2m_2}^{jm}$ of Varshalovich \emph{et al.} \cite{varshalovich88}, which was used by Jahn \cite{jahn51} and Alder \cite{alder52}:

\begin{equation}\label{cle}
C_{j_1m_1,~j_2m_2}^{jm}=(-1)^{j_1-j_2+m}\sqrt{2j+1}\threej{j_1}{j_2}{j}{m_1}{m_2}{-m}.
\end{equation}

\noindent In 1979, Varshalovich and Khersonskii published a simple connection between $\langle r^k\rangle=\langle n\ell|r^k|n\ell\rangle$ and $C_{\ell~n,(k+1)~0}^{\ell~n}$ \cite{varshalovich79}. In 2015, Varshalovich and Karpova found that both $C_{\ell'~n,(k+1)~0}^{\ell~n}$ and $\langle n\ell|r^k|n\ell'\rangle$ can be put in a form in which they are proportional to the hypergeometric function \cite{luke69}:

\begin{equation}
_3F_2\left(\begin{array}{l}
\ell+\ell'-k,\ell-\ell'-k-1,-k-1\\
n+\ell-k,-2k-2\\
\end{array};1\right).
\end{equation}

\noindent The authors obtained the following result \cite{varshalovich15}:

\begin{equation}
\frac{\langle n\ell|r^k|n\ell'\rangle}{C_{\ell'~n,(k+1)~0}^{\ell~n}}=\frac{i^{\Delta}}{2n\sqrt{2\ell+1}}\left(\frac{n}{2}\right)^kf_{\ell,\ell'}^k,
\end{equation}

\noindent where

\begin{equation}
f_{\ell,\ell'}^k=\left[\frac{(k+1-\Delta)!(k+1+\Delta)!(\ell+\ell'+k+2)!}{\left[(k+1)!\right]^2(\ell+\ell'-k-1)!}\right]^{1/2},
\end{equation}

\noindent with $\Delta=\ell-\ell'\geq$ 0, and the Clebsch-Gordan $C_{\ell'~n,(k+1)~0}^{\ell~n}$ is evaluated in a nonphysical region of its arguments, where the so-called triangle condition is not satisfied and the projections of angular momenta $\ell$ and $\ell'$ are larger than the angular momenta themselves. In this region, the Clebsch-Gordan coefficient vanishes, but the product $f_{\ell,\ell'}^k\times C_{\ell'~n,(k+1)~0}^{\ell~n}$ is nonzero. For $(k+1)<0$, coefficient $C_{\ell'~n,(k+1)~0}^{\ell~n}$ transforms into $(-1)^{\ell-\ell'}C_{\ell'~n,-(k+2)~0}^{\ell~n}$ \cite{bandzaitis64} and $f_{\ell,\ell'}^k$ into $f_{\ell,\ell'}^{-k}$. For negative values, one has $(-a)!/(-b)!=(-1)^{b-a}(b+1)!/(a+1)!$.

We have shown recently, {\it mutatis mutandis}, that the expectation values can also be expressed in terms of Clebsch-Gordan coefficients in the relativistic case \cite{pain20}, and we have provided the corresponding relations. In the future, we plan to investigate the case of relativistic non-diagonal elements. 

Group theory has proven to be of great interest for the Coulomb problem, in particular in atomic physics for the understanding of the symmetry properties of the hydrogen atom either in direct or Fock spaces \cite{murnaghan38,decoster70,decoster70a,englefield72,michel16,maclay20}. The application of dynamical groups to atomic physics was reviewed by Wulfman \cite{wulfman79}. In particular, dynamical-group methods have been used to study the symmetry properties of hydrogenic radial wave functions \cite{armstrong70a,armstrong70b,armstrong71}. The basic procedure introduced by Armstrong is to study the $O(2,1)\times O(3)$ subgroup of the $O(4,2)$ dynamical group of the hydrogen atom. Analysis of matrix elements of $r^p$ and $r^{-p}$ shows that the Wigner-Eckart theorem is valid for this group and that the relevant Clebsch-Gordan coefficients are proportional to the usual $SO(3)$ Clebsch-Gordan coefficients. One of the consequences of such a property is an explanation of the Pasternack-Sternheimer selection rule \cite{pasternack62}:

\begin{equation}
\langle n\ell|r^ {-k}|n\ell'\rangle=0\;\;\;\;\mathrm{for}\;\;\;\;(k=2, 3, \cdots, |\ell'-\ell|+1).
\end{equation}

It turns out that the coupling coefficients of the non-compact group $O(2,1)$ are very similar to those of the three dimensional rotation group $SO(3)$ \cite{condon80}. Therefore, the matrix elements of $r^{-k}$, for $k\leq \ell'+\ell-2$ and for varying $n$, can be expressed as

\begin{equation}
\langle n\ell'|\frac{1}{r^k}|n\ell\rangle=(-1)^{\ell'-n}\frac{\mathcal{F}(\ell',\ell,k)}{n^{k+1}}\threej{\ell'}{(k-2)}{\ell}{-n}{0}{n},
\end{equation}

\noindent where $\mathcal{F}$ is a function of $\ell$, $\ell'$ and $k$, but not of $n$. Then the abovementioned selection rule corresponds to a violation of a triangular condition on the triad $(\ell',k-2,\ell)$. For the special case $\ell'=\ell+1$, one has

\begin{equation}
\langle n(\ell+1)|\frac{1}{r^k}|n\ell\rangle=\langle (\ell+1)|\frac{1}{r^k}|\ell\rangle=(-1)^{\ell+1-n}\frac{\mathcal{F}(\ell+1,\ell,k)}{n^{k+1}}\threej{(\ell+1)}{(k-2)}{\ell}{-n}{0}{n}.
\end{equation}

\noindent For instance, for $k=4$, the expectation value

\begin{equation}
\langle n\ell|\frac{1}{r^4}|n\ell\rangle=\frac{Z^4}{2n^5}\frac{3n^2-\ell(\ell+1)}{\left(\ell+\frac{3}{2}\right)(\ell+1)\left(\ell+\frac{1}{2}\right)\ell\left(\ell-\frac{1}{2}\right)}
\end{equation}

\noindent is proportional to

\begin{equation}
\threej{\ell}{2}{\ell}{-n}{0}{n}.
\end{equation}

The fact that radial properties can be cast in this form is highly suggestive \cite{heim09}, and probably opens the way to further interesting developments in atomic shell theory \cite{judd71}. Such properties of the radial function are not well known; this emphasizes the lack of interest which has traditionally been shown in this part of the total wavefunction. Atomic physicists have mostly concentrated their attention on the angular portion of the wavefunction: the spherical harmonics. Group theory has been intensively applied to this study of angular wavefunctions \cite{judd63}. High-energy theoreticians, on the other hand, have been interested in the group properties of the entire wavefunction of the hydrogen atom, and have considered this atom in the light of groups such as $O(4)$, $O(4,1)$ and $O(4,2)$ \cite{barut67a,barut67b,swamy70,musto66,pratt66}.

Armstrong's discussion was concerned with functions of the radial variable $r$, but hydrogenic functions depend in fact on the reduced variable $r/n$. This is the reason why he limited his investigation to matrix elements between states of the same shell. However, it is worth mentioning that Barut and Kleinert \cite{barut67a,barut67b} have considered an algebra which actually has the hydrogenic radial wavefunctions as a basis, and which is therefeore suitable for the study of non-diagonal (with respect to principal quantum number) matrix elements. Unfortunately, powers of $r$ do not transform so conveniently within this algebra.

The application of group theoretical techniques to many-electron atoms is more difficult, although some possibilities exist. Many properties of atoms can be reasonably well predicted using the hydrogenic functions as an approximation to the real wavefunctions, using screened hydrogenic charges for instance. Another possibility would be to search for an approximate potential such that the radial equation could be written in terms of a Casimir operator for some group. The same general approach could then be carried out for the group of the the familiar $SO(3)$ Clebsch-Gordan coefficients in this new potential.

\end{document}